\documentclass[aps,prl,twocolumn,showpacs,floatfix,superscriptaddress]{revtex4-1}

\usepackage{graphicx,color}
\usepackage{amsfonts}
\usepackage[figuresright]{rotating}  
\usepackage{amssymb}
\usepackage{amsmath}

\usepackage{bbold}
\usepackage{mathtools}
\usepackage{psfrag}
\usepackage{subfigure}
\usepackage{multirow}
\usepackage{tabularx}
\usepackage{textcomp}
\usepackage{bm}
\usepackage{hyperref}
\usepackage{relsize}
\hypersetup{
 pdfnewwindow=true, colorlinks=true,
 linkcolor=blue, anchorcolor=blue,
 citecolor=blue, filecolor=blue,
 menucolor=blue, urlcolor=blue}

\def\nn{\nonumber}

\def\beq{\begin{eqnarray}}
\def\eeq{\end{eqnarray}}

\renewcommand{\v}[1]{\ensuremath{\mathbf{#1}}} 
 
\let\baraccent=\= 
\renewcommand{\=}[1]{\stackrel{#1}{=}} 

\makeatletter

\begin{document}

\title{Jerk current: A novel bulk photovoltaic effect}

\author{Benjamin\ \surname{M. Fregoso}}
\affiliation{Department of Physics, Kent State University, Kent, Ohio 44240, USA}

\author{Rodrigo\ \surname{A. Muniz}}
\affiliation{Department of Electrical Engineering and Computer Science, University of Michigan, Ann Arbor, Michigan 48109 USA}
\affiliation{Department of Physics, University of Toronto, Toronto, Ontario, MS5S 1A7, CA}

\author{J.\ \surname{E. Sipe}}
\affiliation{Department of Physics, University of Toronto, Toronto, Ontario, MS5S 1A7, CA}

\begin{abstract}
We investigate a physical divergence of the third order polarization susceptibility representing a photoinduced current in biased crystalline insulators. This current grows quadratically with illumination time in the absence of momentum relaxation and saturation; we refer to it as the \textit{jerk current}. Two contributions to the current are identified. The first is a hydrodynamic acceleration of optically injected carriers by the static electric field, and the second is the change in the carrier injection rate in the presence of the static electric field. The jerk current can have a component perpendicular to the static field, a feature not captured by standard hydrodynamic descriptions of carriers in electric fields. We suggest an experiment to detect the jerk current and some of its interesting features.
\end{abstract}

\maketitle

\textit{Introduction}.-- The dynamics of electrons in electric and magnetic fields gives rise to many novel condensed matter phenomena. Bloch electrons in a static electric field, e.g., acquire an ``anomalous" contribution to their Bloch velocity which leads to the anomalous Hall effect of ferroelectric metals~\cite{Karplus1954}. For slow and nearly homogeneous fields the semiclassical equations of motion provide an accurate description of electron dynamics in metals~\cite{Xiao2010}. Insulators, on the other hand, lack a Fermi surface and a static electric field alone will not generate a steady state current. 

A large dc current can be generated in an insulator by an optical field alone, e.g., via the so-called bulk photovoltaic effect. Its name derives from the fact that there is no need for barriers or interfaces, such as $pn$ junctions, to generate it. Injection~\cite{Sturman1992} and shift currents~\cite{Baltz1981} are examples of the bulk photovoltaic effect. Intuitively, injection current is generated by the asymmetry in carrier injection at time-reversed crystal momenta in the Brillouin zone (BZ). Shift current, on the other hand, arises from the displacement of charge in real space during the process of photon absorption. Both of these are second order effects in the optical electric field and can also be understood as physical divergences of the second order electric polarization susceptibility~\cite{Sipe2000}.

The situation when both an static electric field and optical field are simultaneously present has received less attention. The description of photocurents when both a static and an optical field are present~\cite{Kadanoff1994,Jepsen1996} usually involves the assumption that the carrier injection is solely due to the optical field, e.g., given by Fermi's golden rule, and that injected carriers are then accelerated by the static electric field. However, this description is not complete, as the carrier injection rate itself is affected by the static field. 

In this Letter we reexamine photocurrents in the presence of static field from the perspective of nonlinear optical susceptibilities~\cite{Sipe2000}. This approach starts from a microscopic model of noninteracting Bloch electrons perturbed by an electric field.  By studying the leading divergence of $\chi_3$ (to be defined below), we find a dc current that grows quadratically in time, in the absence of momentum relaxation. We dub it the \textit{jerk current}, by analogy with the use of the word ``jerk" to describe the second derivative of the velocity of a classical particle. We also provide a general expression in terms of microscopic parameters of the material which is suitable for \textit{ab initio} numerical calculations. 

To make contact with the semiclassical description we rederive the same results from a simpler phenomenological model of electron wave packets interacting with light in a static electric field. This perspective allows  us to identify two physically distinct contributions. One arises from the carrier's acceleration, as in the semiclassical description, and a second which arises from the change of the charge carrier injection rate in the presence of a static electric field. This last contribution is completely absent in the usual semiclassical approach.

The jerk current has key differences with other photocurrents studied previously. First, for time scales shorter than the momentum relaxation time, and in the absence of saturation effects, the jerk current grows quadratically with time of illumination, in contrast to the injection current, which grows linearly with time, and the shift current, which is constant. Second, and more important, the jerk current can yield a current perpendicular to the static field. In metals, the Berry curvature of the Fermi surface can lead to a transverse conductivity, e.g., in the anomalous Hall effect, but in insulators without nontrivial Berry curvature~\cite{Morimoto2016} such transverse conductivity is highly unusual. In the standard semiclassical approach the current would flow parallel to the electric field~\cite{Jepsen1996}.

\textit{Physical divergence in susceptibility}. The jerk current is described by a divergence in the third order susceptibility $\chi_3^{abcd}(-\omega_{\Sigma},\omega_{\beta},\omega_{\sigma},\omega_{\Delta})$ when one of the external frequencies is zero. Superscripts indicate Cartesian components, $\omega_{\beta},\omega_{\sigma}$ and $\omega_{\Delta}$ the frequencies of external electric fields, and $\omega_{\Sigma}$ is the sum of external frequencies. The optical electric field is taken to be uniform, $E^b(t)=E^b_{\omega} e^{-i\omega t} + E^{b}_{-\omega} e^{i\omega t}$. Summation over repeated Cartesian indices will be implied. Both the intraband ($\chi_{3i}$) and interband parts of $\chi_3$ have been calculated before for free Bloch electrons~\cite{Aversa1995}. In the limit where $\omega_{\Sigma}$ vanishes, $\omega_{\beta}=\omega,~\omega_{\sigma}=-\omega$, and $\omega_{\Delta}=0$ (static field) we obtain (see Supplementary Material)
\begin{align}
\textrm{lim}_{\omega_{\Sigma}\to 0}~(-i\omega_{\Sigma})^3\epsilon_0 \chi_{3i} = \iota_3,
\label{eq:taylor_chi3}
\end{align}
where $\iota_3$ is finite and vanishes for photon energies smaller than the energy gap; $\epsilon_0$ is the permittivity of free space, and we use a notation for $\chi_3$ and $\iota_3$ which agrees with the standard notation of susceptibilities~\cite{Boyd2008}. Since the current density and macroscopic polarization are related by 
\begin{align}
\frac{d P^a}{d t} &=J^a,
\label{eq:dP_J}
\end{align} 
where $J^a$ is the current density and $P^a$ the macroscopic polarization of the insulator, $\iota_3$ identifies a current given by
\begin{align}
\frac{d^2 J^{a(3)}_{\textrm{jerk}}}{dt^2}=6 \iota_{3}^{abcd}(0,\omega,-\omega,0) E^b_{\omega} E^c_{-\omega} E^d_0,
\label{eq:Jacc}
\end{align}
where $E^d_0$ is a static external field, and the factor of $3!=6$ arises because the response coefficients such as $\iota_3$ and $\chi_3$ are defined to be symmetric under a pairwise exchange of frequency and Cartesian components~\cite{Boyd2008}. For times shorter than the momentum relaxation time and neglecting any saturation effects, Eq.~(\ref{eq:Jacc}) implies that the jerk current grows with time of illumination as
\begin{align}
|\v{J}^{(3)}_{jerk}|\sim \iota_3 t^2.
\end{align}
An explicit calculation (see Supplementary Material) of $\iota_3^{abcd} (0,\omega,-\omega,0)$ gives
\begin{align}
\iota_3^{abcd} = \frac{2\pi e^4}{6\hbar^3 V}\sum_{nm\v{k}}& f_{mn} \big[ 2\frac{\partial^2 \omega_{nm}}{\partial k^d \partial k^a} r^b_{nm}r^c_{mn}\nn\\
&+ \frac{\partial{\omega_{nm}}}{\partial k^a}  \frac{\partial }{\partial k^d}(r^b_{nm}r^c_{mn}) \big] \delta(\omega_{nm}-\omega),
\label{eq:jerk}
\end{align}
where $e=-|e|$ is the charge on the electron and $2\pi\hbar$ is Planck's constant. Here $\hbar \omega_{nm}=\hbar \omega_{n}-\hbar\omega_{m}$ is a band energy difference at crystal momentum $\v{k}$, and $f_{nm}=f_n -f_m$ are the occupation differences of bands $n,m$. The usual dipole matrix elements are denoted by $\v{r}_{nm}$ which by definition vanish for $m=n$. The sums run over all bands $n,m$ and the integration is over all the BZ, with $V$ the sample volume.
 
Using $\v{r}_{nm}(-\v{k})=\v{r}_{mn}(\v{k})$, which can be assumed when there is time reversal symmetry in the unperturbed state, it is easy to show that  $[\iota_{3}^{abcd}(0,\omega,-\omega,0)]^{*}= \iota_{3}^{acbd}(0,\omega,-\omega,0)=\iota_{3}^{abcd}(0,-\omega,\omega,0)=\iota_{3}^{acbd}(0,-\omega,\omega,0)$, and hence $\iota_3$ is real and symmetric in the indices $b,c$. Since $\iota_3$ is a four rank tensor it is nonzero in materials with or without center of inversion. Interestingly, tensor components where $a$ is perpendicular to $d$ lead to current perpendicular to the static electric field even in systems with rotational symmetry.

The longitudinal jerk current can be generated in any material by linearly or circularly polarized, and by unpolarized light. The transverse jerk current, on the other hand, can be generated in any material by linearly polarized light. Unpolarized and circularly polarized light do not necessarily generate a transverse jerk. For example, a zinc blende semiconductor grown along the [100] axis illuminated by circularly or unpolarized light will not support a transverse jerk current for fields perpendicular to the growth direction. Also, the helicity of circularly polarized light does not change the jerk current, which is a consequence of the symmetry of the jerk current tensor $\iota^{abcd}_3$ on the ``$bc$" indices.

Two terms contribute to the jerk current. The first term depends on the curvature of the bands or alternatively, the inverse mass tensor at $\v{k}$ points in the BZ. The second term depends on the momentum derivative of the product of transition matrix elements $r_{nm}^b$ and $r_{mn}^c$. In general, both terms contribute to the current parallel and perpendicular to the static field. From the expression shown in Eq.~(\ref{eq:jerk}) the physical origin of these terms, and whether or not they have analogs in the semiclassical picture, is not clear. To reveal the connection, we now construct a simple phenomenological model where the physics becomes more transparent.

\textit{Phenomenological model}.-- Consider an electron wave packet in band $n$ centered at $\v{k}$. The velocity of the wave packet is $v_{n}^a =\partial \omega_{n}/\partial k^a$. In the presence of a static electric field the wave vector of the electron obeys
\begin{align}
\frac{d \hbar\v{k}}{dt} = e \v{E}_0 = -e\frac{\partial \v{A}}{\partial t},
\end{align}
where the vector potential $\v{A}$ is used to describe the static electric field. If the initial value of $\v{k}$ is $\v{k}_0$ then $\v{k} =\v{k}_0 - e\v{A}/\hbar$ gives an expression for the velocity as a function of $\v{A}$. Expanding in powers of $\v{A}$ and taking a time derivative gives (to first order in the electric field)

\begin{align}
\frac{d v_{n}^a}{d t} = \frac{e}{\hbar}\frac{\partial^2 \omega_{n}}{\partial k^a \partial k^d} E^d_0.
\label{eq:dvdt_exp}
\end{align}
This is the initial acceleration right after the field is turned on. For longer times higher order terms will be important. We assume the system is gapped with all valence bands initially filled and all conduction bands empty. Now the current density is given by
\begin{align}
J^{a} = \frac{e}{V}\sum_{n\v{k}} f_n v_{n}^{a},
\label{eq:j_zero}    
\end{align}
and, if there is a static electric field, it vanishes because the dispersion relations are periodic over the BZ. Taking a time derivative we obtain
\begin{align}
\frac{d J^{a}}{dt} = \frac{e}{V}\sum_{n\v{k}} \left( \frac{d f_n}{dt} v_{n}^{a} +  f_n \frac{d v_{n}^{a}}{dt} \right).
\label{eq:dj}    
\end{align}
If a static electric field is present it will accelerate electrons within the same band but will not excite them across the energy gap (i.e., $df_n/dt=0$). Then using Eq.~(\ref{eq:dvdt_exp}) we have
\begin{align}
\frac{d J^{a(1)}}{dt} \bigg|_{dc} = \frac{e^2}{\hbar V}\sum_{n\v{k}} f_n \frac{\partial^2 \omega_{n}}{\partial k^a \partial k^d} E^d_0,  
\end{align}
which vanishes again, because the dispersion relations are periodic. That is, although in this simple picture each electron accelerates, the net change in current density vanishes because the accelerations add to zero. Note that this holds no matter how many terms we would keep in the expansion of Eq.~(\ref{eq:dvdt_exp}), since only more derivatives of $\omega_n$ would arise. Hence, applying a static electric field alone does not lead to a current at this level of description. There will be a polarization induced as the field is turned on and there will be a current associated with the rise of the polarization, but that is not captured by this simple description. In calculating the current, then, if the static field is on for a long time we can simply assume filled bands when an optical field is turned on. An incident optical field will produce electron-hole pairs and a one-photon absorption calculation is appropriate. The state of the system will be of the form
\begin{align}
|\Psi\rangle = |\textrm{GS}\rangle + \sum_{cv\v{k}}\gamma_{cv} a^{\dagger}_{c} a_{v} |\textrm{GS} \rangle+\cdots,
\end{align}
where $n=c,(v)$ runs over the conduction (valence) bands, $|\textrm{GS}\rangle$ is the ground state of the system with all valence bands filled and conduction bands empty, and $\gamma_{cv}$ is the amplitude for the state $c$ to be occupied and $v$ to be empty (at crystal momentum $\v{k}$); $a_{n}$ ($a^{\dagger}_{n}$) is destruction (creation) operator of Bloch electrons in band $n$. The standard Fermi's golden rule calculation gives the transition rate as
\begin{align}
\frac{d|\gamma_{cv}|^2}{dt} &= 2\pi |\frac{ie}{\hbar \omega}v_{cv}^b E^b_{\omega} |^2\delta(\omega_{cv}-\omega)\\
&= \frac{2\pi e^2}{\hbar^2} r_{vc}^{b} r_{vc}^{c} E^b_{\omega}E^{c}_{-\omega}\delta(\omega_{cv}-\omega),
\label{eq:FGrule}
\end{align}
where we used standard identities $r_{cv}^a = v_{cv}^a/i\omega_{cv}$ and $[v^{a}_{cv}(\v{k})]^{*}=v^a_{vc}(\v{k})$. Alternatively, if we let $f_{v}$ be the occupation of the valence band $v$ and $f_{c}$ the occupation of the conduction band $c$, then their rate of change is
\begin{align}
\frac{d f_{c}}{dt} &= \sum_v \frac{d |\gamma_{cv}|^2}{dt},\nn\\
\frac{d f_{v}}{dt} &= -\sum_c \frac{d |\gamma_{cv}|^2}{dt}. 
\label{eq:df-dt}
\end{align}
If there were no static electric electric applied we have $dv^a_n/dt=0$, and from Eqs.~(\ref{eq:dj}) and (\ref{eq:df-dt}) we recover
\begin{align}
\frac{d J^{a(2)}}{dt}\bigg|_{op} = \frac{2\pi e^3}{\hbar^2 V}\sum_{cv\v{k}} (v_n^a-v^a_m) r^{b}_{vc} r^{c}_{cv} \delta(\omega_{cv}-\omega)E^b_{\omega}E^{c}_{-\omega},
\label{eq:inj_current}
\end{align}
which is the standard injection current expression~\cite{Sipe2000}. The injection current will vanish if the crystal is centrosymmetric because it is governed by a third rank tensor. But now let us assume there is a dc field on when the laser pulse arrives. We assume that over the time we use Fermi's golden rule the static electric field drives the carriers a small fraction of the distance across the BZ within their bands. Then we can use Eq.~(\ref{eq:dvdt_exp}) for each electron. Over the time of the pulse, the velocity of each electron will change, but the sum of all velocity changes cancels out. That need not hold true when we move electrons from one band to another. To see this, take the time derivative of Eq.~(\ref{eq:dj}) to obtain
\begin{align}
\frac{d^2 J^{a}}{dt^2} = \frac{e}{V}\sum_{n\v{k}} \left( \frac{d^2 f_n}{dt^2} v_{n}^{a} +  2\frac{d f_n}{dt} \frac{dv_{n}^{a}}{dt}
+ f_n \frac{d^2 v_{n}^{a}}{dt^2} \right).
\label{eq:d2j}    
\end{align}
One can show that the last term is of second order in $E^d_0$. Then, using Eqs.~(\ref{eq:dvdt_exp}) and (\ref{eq:df-dt}) we have, to linear order in $E_0^d$
\begin{align}
\frac{d^2 J^{a(3)}}{d t^2} = \frac{2\pi e^4}{\hbar^3 V}\sum_{cv\v{k}}2 \frac{\partial \omega_{cv}}{\partial k^d \partial k^a } r_{vc}^b r_{cv}^c \delta(\omega_{cv}-\omega) E^b_{\omega} E^c_{-\omega} E_0^d \nn\\
+\frac{2\pi e^4}{\hbar^3 V}\sum_{cv\v{k}}\frac{\partial \omega_{cv}}{\partial k^a} \frac{\partial (r_{vc}^b r_{cv}^c)}{\partial k^d} \delta(\omega_{cv}-\omega) E^b_{\omega} E^c_{-\omega} E_0^d,
\label{eq:pheno_jerk}
\end{align}
in agreement with Eq.~(\ref{eq:jerk}). An important point of this calculation is to show that the second term in Eq.~(\ref{eq:jerk}) comes from changes in the transition rate $d^2f_n/dt^2$ due to a static field. This contribution is not captured by the semiclassical approach~\cite{Jepsen1996}. The term proportional to $d v_n/dt$ in Eq.~(\ref{eq:d2j}) is the classical acceleration of carriers in the presence of the static field and gives rise to the first term in Eq.~(\ref{eq:jerk}).

\begin{figure}
\includegraphics[width=.45\textwidth]{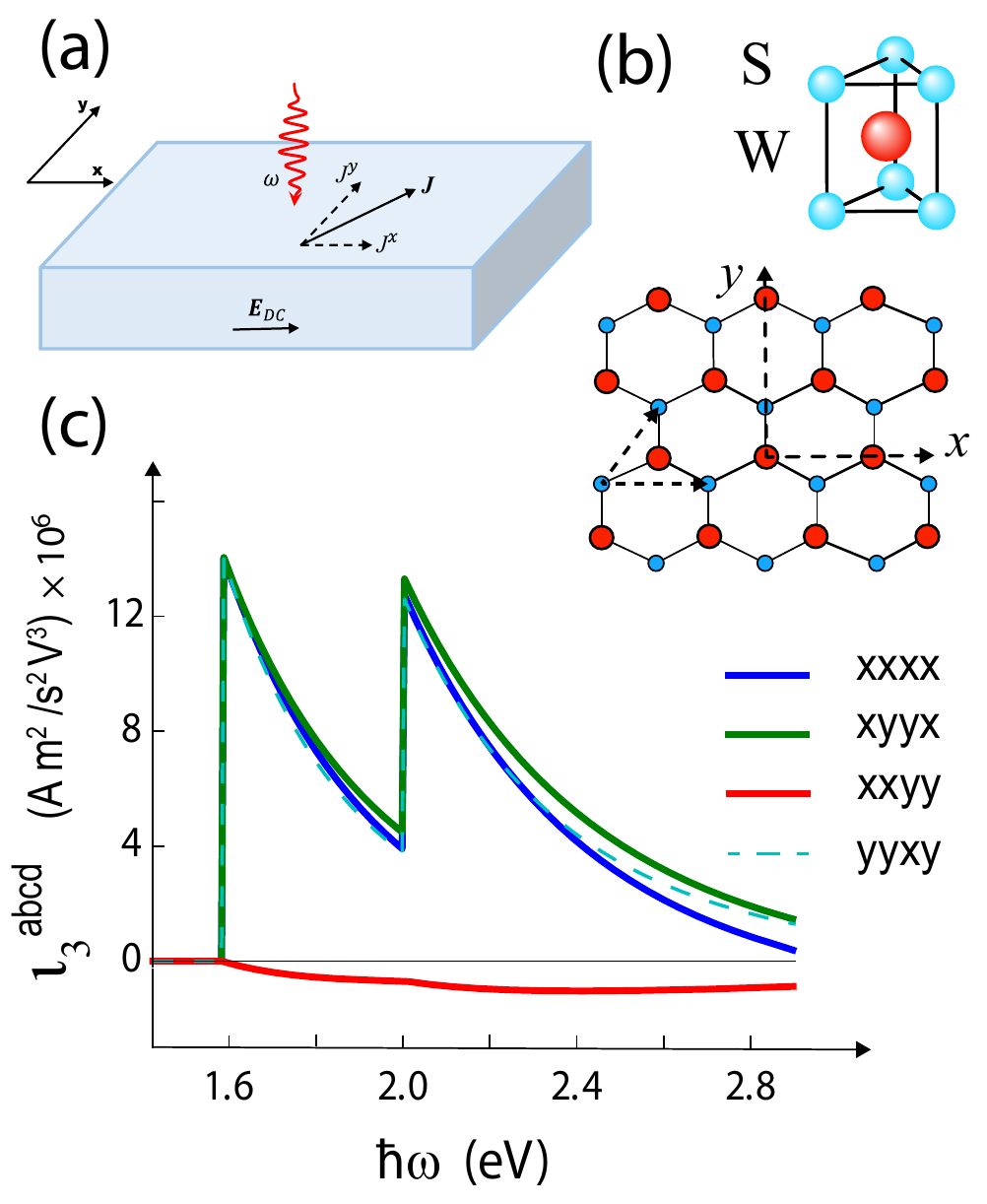}
\caption{(a) Setup for measuring jerk current parallel and perpendicular to the static electric field. (b) The unit cell of single-layer transition metal dichalcogenide WS$_2$ and its $xy$ projection. (c) Jerk current response coefficients for WS$_2$ near the band edges. The model is shown in Eq.~(\ref{eq:H_tmd}). Independent components are shown per valley. There is a nonzero current perpendicular to the static field given by the $xxyy$ component which is about an order of magnitude smaller that the other components. The dashed line indicates the contribution to the $yyxy$ component from the $v=1$ valley; the contribution to this component from the $v=-1$ valley is of the opposite sign.}
\label{fig:jerk_tmd}  
\end{figure}

\textit{An example}.-- A key signature of the jerk current is that it can have a photocurrent perpendicular to the static electric field. We now estimate the order of magnitude of the jerk current perpendicular to the static field in a standard model of single-layer transition metal dichalcogenides~\cite{Xiao2012} (TMDs). The model describes the carriers near the band edges where they behave as gapped Dirac fermions constrained to move in two valleys $v=\pm 1$ (not to be confused with valence band index) for each spin component $s=\pm 1$. The Bloch Hamiltonian for TMDs has four bands and is given by
\begin{align}
H_{v s} = \frac{\hbar}{2} \lambda v s \sigma_0 + \hbar \gamma (v k_x \sigma_x + k_y\sigma_y) + \frac{\hbar}{2} (\Delta-\lambda vs)\sigma_{z}
\label{eq:H_tmd},
\end{align}  
where $\sigma_i$, $i=x,y,z$ are usual Pauli spin matrices and $\sigma_0$ is the unit matrix. The parameters $\gamma,\Delta$, and $\lambda$~\cite{Xiao2012} for MoS$_2$ are $\hbar\gamma=3.5$ {\AA}eV, $\hbar\lambda=0.075$ eV, $\hbar\Delta=1.7$ eV and for WS$_2$, $\hbar\gamma=4.38$ {\AA}eV, $\hbar\lambda=0.21$ eV, $\hbar\Delta=1.79$ eV. In Fig.~\ref{fig:jerk_tmd} we have plotted the independent components of $\iota_3$ for WS$_2$ per valley (summed over spin).  

The tensor component $xxxx$ and $xyyx$ (same first and last Cartesian components) describe the standard current generated parallel to a static electric field. These tensor components are, in general, of equal order of magnitude. A novel feature is the generation of current transverse to the static electric field, which is described by the tensor component $xxyy$ ($=xyxy$), and arises when the incident light is linearly polarized with finite components both parallel and perpendicular to the direction of the static electric field. The tensor component $yyxy$ vanishes when the contributions from all the valleys are included.  However, the contribution from each valley is nonzero, and in Fig.~\ref{fig:jerk_tmd} we plot the contribution from the $v=1$ valley.  With circularly polarized incident light, we would inject a pure valley current, which could be observed with spatially resolved circularly polarized probes.

The jerk current could also be measured in transport experiments, by terahertz spectroscopy or in pump probe experiments. These techniques have been used previously to measure injection and shift currents in various materials~\cite{Priyadarshi2009,Atanasov1996,Laman1999,Bieler2005,Bas2015,Tan2016,Salazar2016}. Here, we propose a terahertz experiment where an insulator is placed in a strong electric field and current is measured perpendicular to the static field. The direction of the jerk current can be inferred from the polarization of the terahertz radiation.  

To estimate the actual current generated we adopt a simple relaxation-time approximation. Let a pulse of duration $\tau\sim 100$ fs and amplitude $E^x_{\omega}=E^y_{\omega}=10^7/\sqrt{2}$ V/m be incident of a sample with static electric field along the $x$-axis of magnitude $E_0= 10^6$ V/m.  Then the photocurrents parallel and perpendicular to the static field are estimated to be  $J_{\textrm{jerk}}^x \approx(\iota^{xxxx}_{3}+\iota^{xyyx}_{3}) \tau^2 |E_{\omega}|^2 E_0\approx 55 ~A/m$ and $J_{\textrm{jerk}}^y \approx (\iota_{3}^{yyxx}+\iota_{3}^{yxyx}) \tau^2 |E_{\omega}|^2 E_0\approx 8 ~A/m$ which is within experimental reach~\cite{Atanasov1996}.

\textit{Conclusions}.-- We have predicted the existence of a novel photocurrent in insulators. It grows quadratically with illumination time until the onset of momentum relaxation or saturation. We showed that the origin of this time dependence has two physical contributions. One is the acceleration carriers in a static electric field, a phenomenon captured by the hydrodynamic description of carriers in a static field. The second is the change in the carrier injection rate in the presence of a static electric field. They are both associated with the physical divergence of the third-order nonlinear susceptibility $\chi_3$. The susceptibility approach to this problem allowed us to give a general expression for the jerk current in terms of material parameters which is also suitable for \textit{ab initio} calculations. 

We have pointed out that the jerk current has a component transverse to the static electric field, a property that is not captured by the hydrodynamic equations of motion of carriers in an electric field. Such an effect could be readily measured in, e.g., TMD and we suggested an experiment to do so. Besides introducing a novel optical effect, our results point out a strategy for controlling currents via optical and static electric fields. 

We acknowledge support from DOE-NERSC Contract No. DE-AC02-05CH11231, and from the Natural Sciences and Engineering Research Council of Canada.


%

\begin{widetext}
\section{Supplementary Material}

\subsection{Derivation of jerk current from the third order susceptibility divergence}
The third order susceptibility $\chi_3^{abcd}(-\omega_{\Sigma},\omega_{\beta},\omega_{\sigma},\omega_{\Delta})$ for free Bloch fermions has been calculated before~\cite{Aversa1995}. For completeness we reproduce the first few terms of the intraband part of $\chi_3$
\begin{align}
\epsilon_0 \chi_{3i}= \frac{e^4}{\omega_2 \omega^2_{\Sigma}\hbar V}\sum_{nm\v{k}} (v^a_n - v^a_m) \left(\frac{r^b_{nm}r^c_{mn} f_{mn}}{\omega_{nm}-\omega_{\beta}} \right)_{;d} 
-\frac{e^4}{\omega^2_{\Sigma}\hbar V}\sum_{nm\v{k}} \frac{(v^a_n - v^a_m)r^d_{mn}}{\omega_{nm}-\omega_{\beta}+\omega_{\sigma}}\left(\frac{r^b_{nm}f_{mn}}{\omega_{nm}-\omega_{\beta}} \right)_{;c} +\cdots,
\label{eq:chi3i}
\end{align}
where $v^a_n=\partial \omega_{n}/\partial k^a$, $\omega_2 =\omega_{\beta}+\omega_{\sigma}$, the electric field is parametrized as $E^b(t)= E^{b}_{\omega}e^{-i\omega t} + E^{b}_{-\omega}e^{i\omega t}$, and various other quantities have already been defined in the main text. The covariant derivative of a matrix element $O^{a}_{nm}$ of Bloch bands $n,m$ is
\begin{align}
O^{a}_{nm;b} =\frac{\partial O^{a}_{nm}}{\partial k^b} - i (\xi_{nn}^b -\xi_{mm}^b )O_{nm}^{a},
\end{align}
where $\xi_{nn}^a = i\langle u_n|\nabla^{a}| u_n\rangle$ are the Berry connections and $u_{n}$ the periodic part of the Bloch state of band $n$. The standard dipole matrix elements are given by $r_{nm}^a= \xi_{nm}^a$ for $n\neq m$ and vanish for $n=m$. We also defined the sum of frequencies as $\omega_{\Sigma} = \omega_{\beta}+\omega_{\sigma}+\omega_{\Delta}$. Eq.~\ref{eq:chi3i} needs to be symmetrized under pair-wise exchange of any two electric field indices~\cite{Boyd2008} $E^{b}_\beta \leftrightarrow E^c_{\sigma} \leftrightarrow E^d_\Delta$. We will show that the first two terms give rise to $\iota_3$. Let us label each term as $\chi_{3i 1}$, and $\chi_{3i 2}$ and consider them separately. By power counting the frequencies in denominators, it is clear that the first term in Eq.~\ref{eq:chi3i} could lead to a finite $\iota_3$ but only when one of the frequencies vanishes.  Without loss of generality let $\omega_{\Delta}= 0$ (and hence $\omega_\Sigma=\omega_2$). Integrating by parts $\chi_{3i1}$, symmetrizing and multiplying by $(-i\omega_{\Sigma})^3$ gives, to zeroth order in $\omega_{\Sigma}$
\begin{align}
(-i\omega_{\Sigma})^3\epsilon_0 \chi_{3i1} =\frac{-i e^4}{6\hbar^3V}\sum_{nm\v{k}} f_{mn} \frac{\partial^2 \omega_{nm}}{\partial k^d \partial k^a} \left( \frac{r^b_{nm} r^c_{mn}}{\omega_{nm}-\omega_{\beta}} +\frac{r^c_{nm} r^b_{mn}}{\omega_{nm}-\omega_{\sigma}} \right). 
\end{align}
Now set
\begin{align}
\omega_\beta &= \omega + n_{\beta}\omega_{\Sigma} \\
\omega_\sigma &= -\omega + n_{\sigma}\omega_{\Sigma},
\end{align}  
$1=n_\beta+n_{\sigma}$, use $1/(x-i\epsilon) =1/x + i\pi\delta(x)$, and expand in powers of $\omega_{\Sigma}$. To zero-order all non-resonant contributions cancel and we obtain 
\begin{align}
(-i\omega_{\Sigma})^3\epsilon_0 \chi_{3i1} =\frac{2\pi e^4}{6\hbar^3 V}\sum_{nm\v{k}} \frac{\partial^2 \omega_{nm}}{\partial k^d \partial k^a}~ f_{mn} r^b_{nm}r^c_{mn}\delta(\omega_{nm}-\omega),
\label{eq:j1}
\end{align}
which contributes to $\iota_3$. There is also a contribution from $\chi_{3i2}$ which we now consider. After symmetrization and integration by parts $\chi_{3i2}$ has 8 terms
\begin{align}
(-i\omega_{\Sigma})^3\frac{\epsilon_0 \chi_{3i2}}{e^4/\hbar^3 V} &=\sum_{l=1}^{8}\chi_{3i2,l}\nn\\
&=-\frac{i\omega_{\Sigma}}{6}\sum_{nm\v{k}} \frac{(v^a_n-v^a_m) r_{mn}^c f_{mn}}{\omega_{nm}-\omega_{\beta}-\omega_{\Delta}} \left(\frac{r_{nm}^b}{\omega_{nm}-\omega_{\beta}} \right)_{;d} 
-\frac{i\omega_{\Sigma}}{6}\sum_{nm\v{k}} \frac{(v^a_n-v^a_m) r_{mn}^b f_{mn}}{\omega_{nm}-\omega_{\sigma}-\omega_{\Delta}} \left(\frac{r_{nm}^c}{\omega_{nm}-\omega_{\sigma}} \right)_{;d} + \cdots
\end{align}
In these two terms the imaginary parts of the frequency play an important role as the reals parts vanish. After an expansion in powers of $\omega_{\Sigma}$ we have to leading order
\begin{align}
\chi_{3i2,1} + \chi_{3i2,2}  = \frac{2\pi}{6}\sum_{nm\v{k}} \frac{\partial}{\partial k^d}(\frac{\partial \omega_{nm}}{\partial k^a} r_{mn}^c r_{nm}^b) f_{mn}\delta(\omega_{nm}-\omega).
\label{eq:j2}
\end{align}
This term also contributes to $\iota_3$. Adding Eq.~\ref{eq:j1} and Eq.~\ref{eq:j2} gives $\iota_3$ in Eq.~\ref{eq:jerk} in the main text. 

\subsection{Jerk current in transition metal dichalcogenides}
The effective Hamiltonian for transition metal dichalcogenides (TMDs) near the band edges is
\begin{align}
H_{v s} = \frac{\hbar}{2} \lambda v s \sigma_0 + \hbar \gamma (v k_x \sigma_x + k_y\sigma_y) + \frac{\hbar}{2} (\Delta-\lambda vs )\sigma_{z}
\label{eq:H_tmd2}.
\end{align}  
It is easy to calculate its eigenvalues and eigenfunctions for each valley ($v$) and spin ($s$) and with those compute the corresponding dipole matrix elements $r_{nm}^a$ and the jerk current. Defining $\Delta_{vs}=(\Delta-\lambda v s)/2$ the results are
\begin{align}
\iota_{3v s}^{xxxx}(0,\omega,-\omega,0)= &- \theta\left(\omega-2\Delta_{v s}\right)\frac{e^{4}\gamma^{2}}{96\hbar^{3}\omega^{2}}\left(10-\frac{116\Delta_{v s}^{2}}{\omega^{2}}+\frac{48\Delta_{v s}^{4}}{\omega^{4}}\right),\\
\iota_{3v s}^{xyyx}(0,\omega,-\omega,0)=&- \theta\left(\omega-2\Delta_{v s}\right)\frac{e^{4}\gamma^{2}}{96\hbar^{3}\omega^{2}}\left(6-\frac{92\Delta_{v s}^{2}}{\omega^{2}}+\frac{16\Delta_{v s}^{4}}{\omega^{4}}\right),\\
\iota_{3v s}^{xxyy}(0,\omega,-\omega,0)=&- \theta\left(\omega-2\Delta_{v s}\right)\frac{e^{4}\gamma^{2}}{48\hbar^{3}\omega^{2}}\left(1-\frac{16\Delta_{v s}^{4}}{\omega^{4}}\right),\\
\iota_{3v s}^{xxyx}(0,\omega,-\omega,0)=&-\theta\left(\omega-2\Delta_{v s}\right)i v \frac{e^{4}\gamma^{2}\Delta_{v s}}{12\hbar^{3}\omega^{3}}\left(1-\frac{28\Delta_{v s}^{2}}{\omega^{2}}-\frac{32\Delta_{v s}^{4}}{\omega^{4}}\right),
\end{align}
The last term is only relevant for circularly polarized light. This term will vanish when summing over valleys. A current perpendicular to $E^d_0$ is possible only for linear polarization along a direction that is neither parallel nor perpendicular to $E^d_0$. 

\end{widetext}

\end{document}